\documentclass[a4paper,fleqn]{cas-sc}

\usepackage[authoryear]{natbib}
\usepackage{graphicx}
\usepackage{amsmath, amssymb}
\usepackage{amsfonts}
\usepackage{enumitem}
\usepackage[most]{tcolorbox}
\usepackage{xcolor}
\usepackage{graphicx}
\usepackage{subcaption}
\usepackage{caption}
\usepackage{float}
\usepackage{booktabs}
\usepackage{multirow}
\usepackage{adjustbox}
\usepackage{tabularx}

\definecolor{promptbg}{RGB}{245,248,250}     
\definecolor{promptframe}{RGB}{0,80,130}     
\definecolor{titlegray}{RGB}{219,238,244}       
\def\tsc#1{\csdef{#1}{\textsc{\lowercase{#1}}\xspace}}
\tsc{WGM}
\tsc{QE}
\tsc{EP}
\tsc{PMS}
\tsc{BEC}
\tsc{DE}

\begin{document}
\let\WriteBookmarks\relax
\def\floatpagepagefraction{1}
\def\textpagefraction{.001}

\shorttitle{An agentic workflow for detecting PII in crash narratives}

\shortauthors{Junyi Ma et~al.}

\title [mode = title]{An Agentic Workflow for Detecting Personally Identifiable Information in Crash Narratives}                      

\author[1]{Junyi Ma}
\ead{jma333@wisc.edu}

\author[2]{Pei Li}
\cormark[1]
\ead{pei.li@uwyo.edu}

\author[1]{Rui Gan}
\ead{rgan6@wisc.edu}

\author[3]{Kai Cheng}
\ead{kc2859@columbia.edu}

\author[1]{Steven T. Parker}
\ead{sparker@engr.wisc.edu}

\author[1]{Bin Ran}
\ead{bran@wisc.edu}


\affiliation[1]{organization={Department of Civil and Environmental Engineering, University of Wisconsin-Madison},
    city={Madison},
    state={WI},
    country={USA}}

\affiliation[2]{organization={Department of Civil and Architecture Engineering and Construction Management, University of Wyoming},
    city={Laramie},
    state={WY},
    country={USA}}

\affiliation[3]{organization={Institute for Social and Economic Research and Policy, Columbia University},
    city={New York},
    state={NY},
    country={USA}}
    
\cortext[cor1]{Corresponding author}

\begin{abstract}
Crash narratives in crash reports provide crucial contextual information for traffic safety analysis. Yet, their broader use is hindered by the presence of personally identifiable information (PII), including names, home addresses, and license plate numbers. Because PII appears sparsely and inconsistently in crash narratives, manual detection is not scalable, and existing rule-based approaches often fail to capture context-dependent PII. This study develops and evaluates a locally deployable, agentic workflow for PII detection in crash narratives by leveraging large language models (LLMs). The workflow contains a Hybrid Extractor and a Verifier. The Hybrid Extractor routes structured PII (e.g., phone numbers and email addresses) to a rule-based model (i.e., Presidio) and context-dependent PII (e.g., names, home addresses, and alphanumeric identifiers) to a domain-adapted, fine-tuned LLM. To address ambiguity in challenging categories, the workflow incorporates ensemble LLM extraction and an agentic verification step that filters false detections through evidence-based reasoning. Evaluated on a real-world crash dataset, the agentic workflow achieves strong performance with a precision of 0.82, a recall of 0.94, an F1 of 0.87, and an accuracy of 0.96, outperforming multiple baseline methods. Moreover, the ablation results suggest that ensemble LLM extraction and Verifier offer improved detection for home addresses and alphanumeric identifiers. The workflow runs locally, supporting privacy-sensitive operational settings where external APIs are restricted. This work offers a practical and robust path for scalable, privacy-preserving crash data processing, enabling broader research and safety interventions while safeguarding individual privacy.
\end{abstract}




\maketitle

\section{Introduction}
Police crash reports are fundamental to safety analysis~\citep{fhwa_hsm_web}. With both structured and unstructured elements, crash reports provide essential information for analyzing crash contributing factors, developing countermeasures, and enhancing safety. The unstructured element of a crash report, the crash narrative, is a detailed description of a crash from the reporting officer. This narrative provides detailed descriptions of specific circumstances and contributing factors that may not be readily available from the structured element \citep{kim2021crash}. As a result, there has been increasing interest in using crash narratives for deeper and more comprehensive safety analysis~\citep{roque2019topic,kwayu2021discovering,lee2023advancing,li2024analyzing}. 

However, crash narratives may contain Personally Identifiable Information (PII), posing risks of exposing personal information and limiting access to crash data. PII refers to any data that can be used to identify an individual, either directly or indirectly. For example, names, home addresses, and phone numbers are usually considered as PII. As a consequence, this valuable dataset is often restricted to internal usage with limited external availability. Identifying such private information, therefore, becomes critical for more accessible safety data.

Most existing studies have focused on analyzing crash narratives rather than enhancing their accessibility~\citep{mumtarin_large_2023,fan2024learning,abdelrahman2025advanced}. To address this limitation, it is important to develop a method that detects PII from crash narratives, making crash data more accessible and available. Given the size of crash datasets and the random appearance of PII, manual identification is not feasible, driving the need for automatic solutions. Although PII detection has been widely studied in various areas, including health care~\citep{uzuner2008identifier}, education~\citep{buchh_enhancing_2024}, and social media~\citep{liu_automated_2021}, very limited attention has been given to transportation. Prior PII detection approaches generally fall into two categories. Rule-based methods are effective for highly structured entities but often overgeneralize or fail in context-sensitive cases. More recent Large Language Models (LLM) approaches offer stronger contextual understanding, but their performance depends heavily on domain alignment, annotation quality, and deployment constraints. In crash narratives, this challenge is especially pronounced because many strings that resemble PII (e.g., street names, roadway identifiers, case numbers) in other domains are not necessarily sensitive under transportation privacy rules. This makes PII detection in crash narratives a domain-specific problem rather than a direct transfer of general identification methods.

In our earlier study~\citep{ma2026crashpii}, we evaluated local, privacy-compatible PII detection approaches for crash narratives, including Microsoft Presidio, prompt-based LLM extraction, and fine-tuned LLM extraction. Those results showed that fine-tuning a local LLM substantially improves performance for context-dependent PII, while rule-based methods remain strong for structured categories. The results also suggested that no single extractor performs best across all PII types. This creates the need for a hybrid framework that leverages strengths from different methods for robust and accurate PII detection.

This paper extends our prior study by introducing an agentic workflow for PII detection in crash narratives. The workflow has a Hybrid Extractor equipped with tools for detecting PII, and a Verifier validates the detection results. Specifically, the Hybrid Extractor uses Presidio for detecting structured PII and a fine-tuned LLM for detecting context-dependent PII. To improve performance on the most ambiguous PII, we employ ensemble learning that pools candidate spans from repeated LLM runs. The Verifier uses a local prompt-engineered LLM to validate detection results, preventing both over-and under-detection. All components of the proposed workflow are deployed locally, which is important for privacy-sensitive environments where large-scale narrative processing must remain within institutional infrastructure. Experimental results on real-world crash narratives indicate that the Hybrid Extractor improves overall category-level performance and that the verification stage further strengthens robustness in categories with high contextual ambiguity. Moreover, comparison results suggest that the proposed workflow outperforms existing baseline methods with more accurate and more robust detection results. 

The contributions of this work are threefold. First, we develop a domain-specific, agentic workflow for PII detection in crash narratives. The workflow is locally deployable, addressing privacy concerns in crash data analysis. Second, we introduce a hybrid extraction design that combines a rule-based method and a fine-tuned LLM for detecting both structured and context-dependent PII. A targeted verifier is integrated in the workflow to address challenges in detecting the most ambiguous PII. Third, we validate the proposed workflow on real-world crash narratives and show that it achieves strong overall performance compared with existing baselines. The remainder of this paper is organized as follows. Section 2 reviews related work. Section 3 describes the proposed method and data. Section 4 reports results and discussions. Section 5 concludes the paper.

\section{Related Work}
Free‑text crash narratives pose a distinct privacy challenge because they mix structured crash variables with open‑ended prose where PII can appear in distinctive ways (e.g., names, phone numbers, license plates, addresses, social security numbers). The U.S. Department of Labor defines PII as any information that can be used to distinguish or trace an individual’s identity, either alone or when combined with other information. This underscores why even seemingly benign fragments in crash narratives matter.

Automated systems for PII extraction have been developed from rule-based and deep learning methodologies. Rule-based methods typically rely on predefined patterns or heuristics to detect various forms of PII. Such techniques have been successfully applied across different textual domains. For example, the authors~\citep{lison-etal-2021-anonymisation,pilan2022text} employed a rule-based approach for named entity recognition and achieved an F1 score of 81\% at the token level in financial documents. Similarly, other researchers~\citep{bosch2020hello} attained up to 95\% recall for identifying student names on educational data from a university. While rule-based systems have shown some effectiveness in handling complex PII detection tasks, their primary limitation lies in limited generalizability. These systems often depend on dataset-specific textual cues that may not transfer well across diverse text sources.

Recent advancements in automated text de-identification have increasingly adopted deep learning methods, particularly those based on transformer architectures. Introduced by Vaswani et al.~\citep{vaswani2017attention}, transformers are neural networks that have become foundational in natural language processing tasks. They currently represent the leading approach for a range of language tasks, including named entity recognition (NER), which involves detecting and categorizing specific entities such as names of people, organizations, or locations. The task of identifying PII can be seen as a specialized application within the broader NER framework. Transformer-based models have demonstrated strong performance in deidentifying medical notes. Researchers achieved a recall of 0.99 and precision of 0.97 on a dataset from Mayo Clinic, by employing a BERT-based cased model~\citep{murugadoss_building_2021}. These models come with notable limitations. They typically demand extensive annotated training data, which must be created through labor-intensive manual labeling, and they require substantial computational resources, posing challenges for settings where data privacy necessitates local processing on low-resource machines. Although transformer-based approaches hold significant promise for handling complex PII detection tasks, their effectiveness has yet to be evaluated in crash narratives.

With a more recent advancement in artificial intelligence, LLMs such as GPT, Llama, and Gemini have shown great potential in contextual text understanding as they are pretrained on an extensive amount of data. For the PII detection task using LLMs, studies exist in  domains such as education data and medical records. One of the first applications of ChatGPT/GPT-4 in medical text de-identification introduces DeID-GPT, a GPT-4-enabled framework for automatically de-identifying unstructured medical text using zero-shot in-context learning~\citep{liu2023deid}. Leveraging the strong NER capabilities of large language models, DeID-GPT outperforms existing methods in accurately masking sensitive information while preserving text structure and meaning. Shen et al. evaluate GPT-4o-mini for PII detection in educational data, comparing prompting and fine-tuning approaches against tools like Microsoft Presidio and Azure AI Language~\citep{shen_enhancing_2025}. Fine-tuned GPT-4o-mini outperforms competitors on public datasets (with 0.9589 recall), while significantly reducing computational costs and maintaining fairness across demographic groups. Researchers also evaluate GPT-4 for de-identifying student-generated content from discussion forums, achieving a high recall of 0.958 in detecting PII~\citep{shreya_singhal_-identifying_2024}. The model identified instances missed by human annotators, highlighting its potential to enhance both the efficiency and quality of the redaction process. However, its lower precision (0.526) indicates a tendency to over-redact, suggesting room for refinement. 

While prompting and fine-tuning improves the generator side of PII extraction, recent work on LLM systems increasingly separates generation from evaluation, using multi-step or role-specific workflows instead of a single-pass response. ReAct-style agents~\citep{yao2022react} formalized the broader idea of structured reasoning / acting loops for LLM agents, highlighting the value of decomposing a task into explicit stages rather than relying on a monolithic generation step. Self-Refine and Reflexion further demonstrated that iterative feedback, critique, and revision can improve output quality without changing the base model weights, by explicitly turning generation into a multi-step refinement process~\citep{madaan2023self,shinn2023reflexion}. This idea is also appearing in applied domain workflows. For example, a recent clinical study on an agentic multi-agent LLM workflow reports that task-specialized agents can achieve strong performance while improving specificity and reducing manual prompt-refinement burden in a high-stakes text classification setting~\citep{tian2025agentic}. Although the task is different from PII extraction, the architectural lesson is directly relevant: decomposing a difficult task into specialized sub-roles can improve robustness and operational reliability.

Taken together, prior work establishes the value of rule-based systems for structured PII, the strength of transformer and LLM methods for contextual entities, and the promise of multi-stage or agentic LLM workflows for reliability. However, to our knowledge, these ideas have not been integrated and evaluated for PII detection in the crash narrative domain under a locally deployable, privacy-preserving setup. This study addresses that gap by evaluating a domain-specific pipeline that combines rule-based extraction, fine-tuned local LLM, and an agentic verification stage.

\section{Methods}
\begin{figure}
    \centering
    \includegraphics[width=1\linewidth]{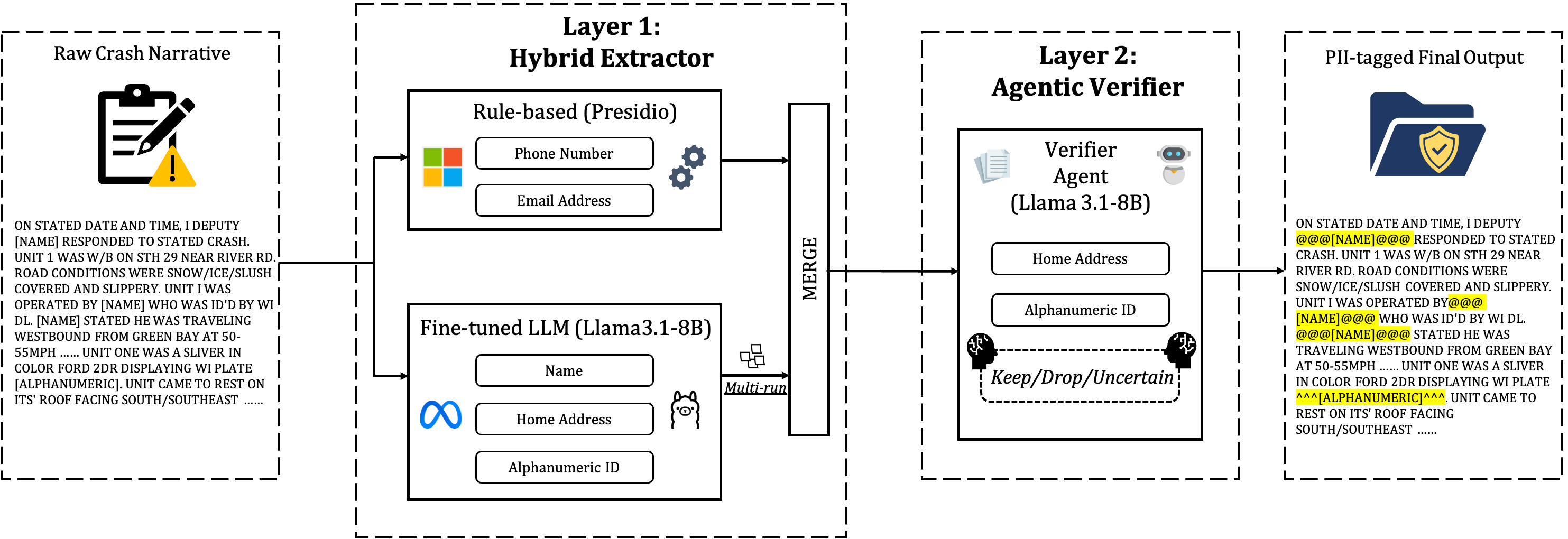}
    \caption{PII detection with the proposed agentic workflow.}
    \label{fig:framework}
\end{figure}

Figure~\ref{fig:framework} illustrates the task definition and our proposed agentic workflow. Given a crash narrative, the two-layer agentic workflow identifies PII and produces a tagged output that preserves non-PII content. 

\subsection{Data Description}
PII refers to any data that can be used to identify an individual either directly or indirectly. While the specific types of PII may vary across domains such as healthcare, education, and finance, they generally include elements like names, home addresses, phone numbers, email addresses, and government-issued identifiers. In transportation, official guidelines often consider the following to be PII: full name, home address, date of birth, phone number, email address, social security number, insurance policy number, driver’s license number, and vehicle license plate number, among others. This study defines the following elements as PII in consideration of federal and state transportation data privacy guidelines~\citep{nhtsa_sgo_crash_reporting,wisconsin_statute_19_62_5}, including:

\begin{enumerate}
    \item {Name}: Full names or initials that can be used to identify individuals involved in the crash.
    \item {Phone number}: Personal or work-related phone numbers included in the narrative.
    \item {Email address}: Any standard email format identifying a specific individual or organization.
    \item {Home address}: Residential addresses, including street names, house numbers, and apartment identifiers. Crash location does not count.
    \item {Alphanumeric identifiers}: Structured combinations of letters and numbers that may include driver’s license numbers, insurance policy numbers, vehicle license plates, or other unique codes.
\end{enumerate}

This definition reflects two practical characteristics of crash narratives. First, some PII types (e.g., phone numbers and email addresses) are relatively structure-dominant and can often be recognized via uniform patterns. Second, others (e.g., names, home addresses and alphanumeric identifiers) are frequently context-dependent. The same text can be non-PII in one context (e.g., roadway identifiers like “I-43”, “US-12”, or location descriptions such as “140TH ST”) but sensitive in another (e.g., “vehicle registered to 140TH ST”). Because of this ambiguity, we treat home address and alphanumeric identifiers as higher-risk categories that require careful review in downstream processing.

This study utilizes crash narrative data provided by the Wisconsin Department of Transportation, covering the years 2017 through 2022. Each year contains approximately 100,000 to 150,000 records, resulting in a substantial dataset for PII detection research. The narrative contains text descriptions written by law enforcement officers, which may include PII about individuals involved in crashes. Our goal is to automatically detect and label PII spans in each narrative so that the identified text can be redacted or replaced with non-sensitive placeholders. 500 narratives were randomly sampled to construct a test dataset, serving as the benchmark for model evaluation. These narratives were then manually reviewed and annotated. Of the 500 narratives, 64 narratives include names, 11 include phone numbers, 3 include email addresses, 16 include alphanumeric characters, and 7 include home addresses.

For model training and evaluation, we operationalize extraction using a span-tagging representation. Each detected PII span is enclosed by a category-specific marker in the output text, enabling straightforward parsing of predicted entities and exact string recovery from the narrative. Specifically, we use the following tags: \texttt{@@@...@@@} for names, \texttt{\&\&\&...\&\&\&} for phone numbers, \texttt{\%\%\%...\%\%\%} for emails, \texttt{\$\$\$...\$\$\$} for home addresses, and \texttt{\textasciicircum\textasciicircum\textasciicircum...\textasciicircum\textasciicircum\textasciicircum} for alphanumeric identifiers. This design is consistent with our prior studies and the literature, which have found that it can reduce model hallucinations and over-labeling~\citep{wang2023gpt}. If a narrative contains no PII, the system returns the input unchanged.

\subsection{Agentic PII Detection Workflow}
This work proposes an agentic PII detection workflow feasible for local deployment. It is designed based on results from our preliminary experiments~\citep{ma2026crashpii}. First, structured PII such as phone numbers and email addresses are consistently detected by rule-based patterns. Second, context-dependent PII including personal names, home addresses, and alphanumeric identifiers often requires semantic cues. Therefore, the key design principle is to combine (i) a high-precision rule-based detector for structured PII with (ii) a detector that understands semantic information for context-dependent PII, and (iii) an agentic verifying step to reduce systematic errors in the most ambiguous categories. 

\subsubsection{\textbf{Layer 1: Hybrid Extractor}}
\textit{Presidio for structured PII}: Phone numbers and email addresses exhibit stable patterns and are well-suited for deterministic detection. We adopt Microsoft Presidio~\citep{microsoft_presidio} as a local rule-based tool for these two PII categories. Presido is an open-source method for identifying and redacting PII in unstructured text.
In our implementation, we utilized Presidio’s default AnalyzerEngine to extract email addresses from crash narratives. Meanwhile specifically for phone number detection, we replace the default recognizer with a stricter U.S. phone pattern to reduce spurious matches and align with formats observed in crash narratives.

\textit{Fine-tuned LLM for contextual PII}:
Names, home addresses, and alphanumeric identifiers frequently require semantic interpretation. For example, street strings may refer to crash locations rather than residences, and alphanumeric patterns may correspond to roadway IDs or internal report numbers rather than personal identifiers. To address these context-heavy categories, we used the Llama 3.1-8B model~\citep{grattafiori2024llama} with both prompt engineering and fine-tuning techniques. The prompt includes explicit role instructions, category definitions, and tagging rules to guide the model’s reasoning process. These techniques have been proven to be effective in helping LLMs obtain domain knowledge with minimal additional effort~\citep{fagbohun2024empirical}. We adopted a tagging-based output format to preserve span alignment and enable exact-string extraction for evaluation as Figure~\ref{fig:llama_prompt_prefix} shows.

\begin{figure*}
\centering
\begin{tcolorbox}[
    width=0.98\textwidth,
    colback=gray!3,
    colframe=black!25,
    boxrule=0.6pt,
    arc=2mm,
    left=2mm,right=2mm,top=1.8mm,bottom=1.8mm
]
{\small
\textbf{Fine-tuning prompt}

\vspace{1mm}

\begin{tcolorbox}[
    colback=white,
    colframe=black!18,
    boxrule=0.4pt,
    arc=1.5mm,
    left=2mm,right=2mm,top=1.2mm,bottom=1.2mm
]
{\small
\textbf{Role:} You are an expert linguist specializing in detecting personally identifiable information (PII) in crash narratives.

\vspace{0.8mm}
\textbf{Context:} Your task is to find and tag any Personally Identifiable Information (PII) using the special identifiers below, based on the PII category. If no PII is found, return the input text unchanged.

\vspace{0.8mm}
\textbf{PII categories and tagging rules:}
\begin{itemize}[leftmargin=5mm,itemsep=0.5mm,topsep=0.5mm]
    \item \textbf{Name}: tag with \texttt{@@@Text@@@} \\
    Example: \texttt{@@@John Smith@@@}

    \item \textbf{Phone Number}: tag with \texttt{\&\&\&Text\&\&\&} \\
    Example: \texttt{\&\&\&608-733-8366\&\&\&}

    \item \textbf{Home Address}: tag with \texttt{\$\$\$Text\$\$\$} \\
    Example: \texttt{\$\$\$123 Elm Street\$\$\$} \\
    Do \textbf{not} tag crash-location addresses. Only tag home addresses based on context.

    \item \textbf{Email Address}: tag with \texttt{\%\%\%Text\%\%\%} \\
    Example: \texttt{\%\%\%jsmith@gmail.com\%\%\%}

    \item \textbf{Alphanumeric Identifiers}, including driver's license, SSN, and license plate number: tag with \texttt{\textasciicircum\textasciicircum\textasciicircum Text\textasciicircum\textasciicircum\textasciicircum} \\
    Example: \texttt{\textasciicircum\textasciicircum\textasciicircum ABC1234\textasciicircum\textasciicircum\textasciicircum}
\end{itemize}

\vspace{0.8mm}
\textbf{The input is:} 
}
\end{tcolorbox}
}
\end{tcolorbox}

\caption{\textbf{Prompt used for fine-tuning LLM.}
The prompt instructs the model to identify candidate PII spans and annotate them with category-specific delimiter markers before downstream verification.}
\label{fig:llama_prompt_prefix}
\end{figure*}

\begin{figure}
   \centering
   \includegraphics[width=0.75\linewidth]{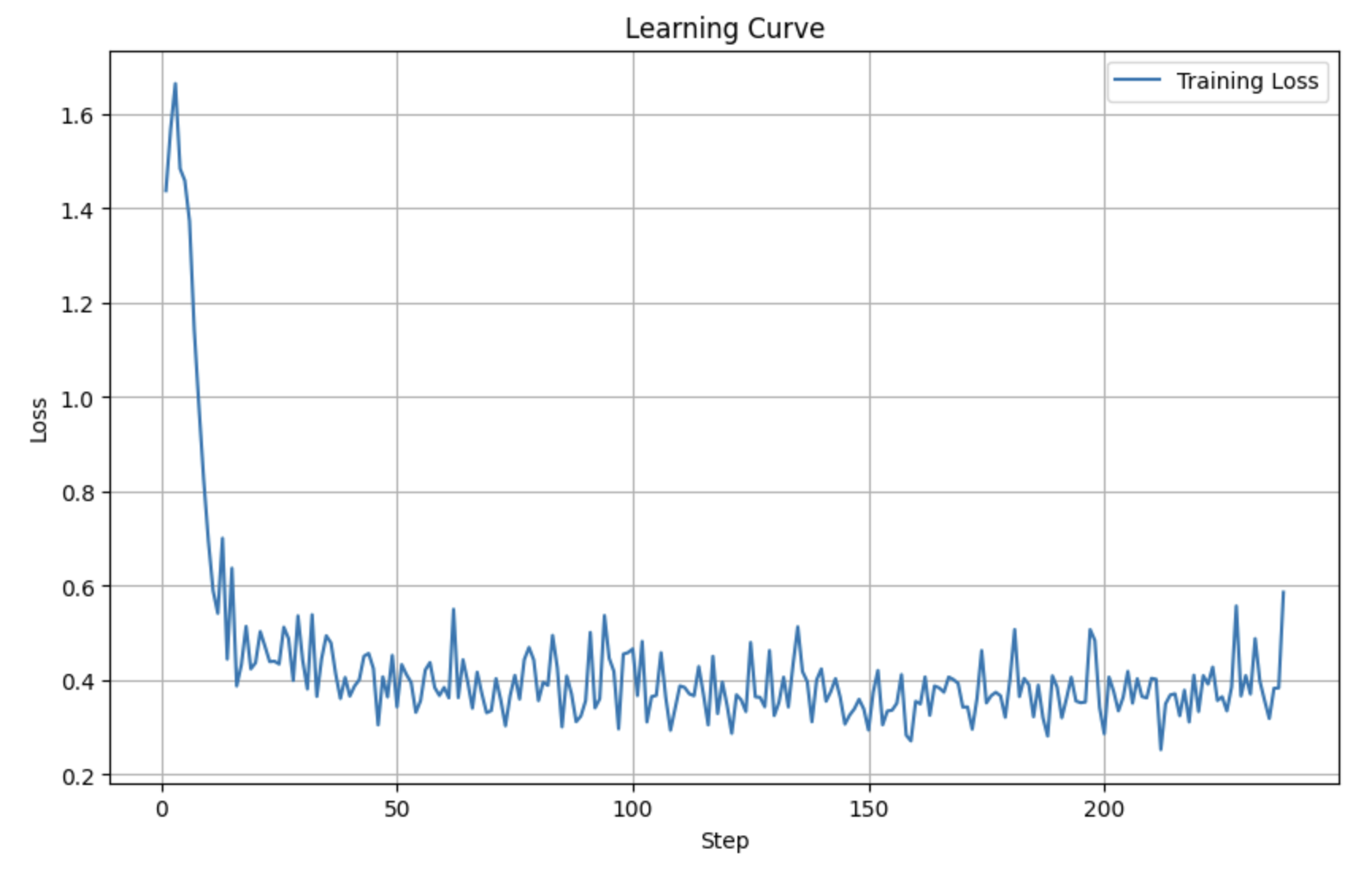}
   \caption{Training curve of the fine-tuned model}
   \label{fig:training curve}
\end{figure}

Although prompt engineering can enhance model performance in detecting PII. The improvement is limited due to the lack of domain knowledge in the LLM. Therefore, we fine-tuned the LLM on manually labeled crash narratives using Low-Rank Adaptation (LoRA)~\citep{hu2022lora}. LoRA is a parameter-efficient technique that enables targeted task adaptation without retraining the entire model. Moreover, early experiments indicated that fine-tuning the LLM on out-of-domain PII datasets~\citep{stubbs2015automated,sang2003introduction} could not meet our task-specific requirements. Not only was it unable to recognize many domain-relevant PII elements (e.g., license plate numbers and home addresses), but more critically, the output format deviated significantly from our requirement. Therefore, we manually annotated 2,000 crash narratives randomly sampled from the dataset using the defined PII categories and specific tagging rules. This manually-labeled dataset provides the domain-relevant supervision necessary for effective fine-tuning~\citep{gan2026crashsightphaseawareinfrastructurecentricvideo}. The fine-tuned Llama 3.1-8B model was trained with 2,000 manually labeled crash narratives for 1 epoch, using a single L4 GPU on Google Colab. As Figure~\ref{fig:training curve} shows, the loss gradually decreased and eventually converged, indicating that the model was able to detect PII from the training data.

\textit{LLMs with ensemble learning}: In practice, the fine-tuned LLM occasionally produces slightly different outputs across runs depending on decoding randomness. To increase recall without changing model weights, we perform an ensemble extraction step. The fine-tuned tagger is executed K times (e.g., K=5), and the union of extracted home addresses and alphanumeric candidates is collected. 

\textit{Hybrid Extractor}: Outputs from Presidio and the fine-tuned LLM are combined into a unified candidate set following a fixed responsibility split. Specifically, Presidio detects phone numbers and email addresses, while the fine-tuned LLM identifies names, home addresses, and alphanumeric identifiers. This split reflects the strengths observed in our experiments: rule-based extraction is stable for highly structured entities, while the fine-tuned model better captures context-dependent entities. The proposed Hybrid Extractor is an orchestration design rather than a new standalone extraction model, it combines complementary extractors under the fixed responsibility split. 

\subsubsection{\textbf{Layer 2: Agentic Verifier}}
Although the Hybrid Extractor improves overall coverage, home addresses and alphanumeric identifiers remain challenging to detect due to extra contextual ambiguity. To address this, we introduce a verifier agent that reviews these two types of PII produced by the extractor. 

The verifier is a locally deployed Llama 3.1-8B that receives the raw narrative and the detected home addresses and alphanumeric identifiers. As Figure \ref{fig:verifier_system_prompt} shows, the model is prompted as a strict extraction reviewer and produces structured decisions for each candidate item. To prevent hallucinations and enforce traceability, the verifier is required to provide a short verbatim evidence snippet from the narrative whenever it outputs KEEP or DROP. This mirrors the role separation commonly used in agentic workflows, where a “specialist” generates an output and a separate “verifier” validates it according to strict criteria. In addition, we enforce strict output formatting rules so that the verifier produces exactly one decision per candidate and does not invent new candidates. 

\begin{figure*}
\centering
\begin{tcolorbox}[
    width=0.98\textwidth,
    colback=gray!3,
    colframe=black!25,
    boxrule=0.6pt,
    arc=2mm,
    left=2mm,right=2mm,top=1.8mm,bottom=1.8mm
]
{\small
\textbf{Verifier system prompt}

\vspace{1mm}

\begin{tcolorbox}[
    colback=white,
    colframe=black!18,
    boxrule=0.4pt,
    arc=1.5mm,
    left=2mm,right=2mm,top=1.2mm,bottom=1.2mm
]
{\small
\textbf{Role:} You are a strict PII extraction verifier for crash narratives.

\vspace{0.8mm}
\textbf{Context:} You will be given:
\begin{itemize}[leftmargin=5mm,itemsep=0.2mm,topsep=0.5mm]
    \item the raw narrative;
    \item extracted candidates for \texttt{HOME ADDRESS} and \texttt{ALPHANUMERIC IDENTIFIERS}.
\end{itemize}

\vspace{0.8mm}
Your job is to decide for each provided candidate: KEEP, DROP, or UNCERTAIN.

\vspace{0.8mm}
\textbf{Critical output format rules (must follow exactly):}
\begin{itemize}[leftmargin=5mm,itemsep=0.35mm,topsep=0.5mm]
    \item \texttt{home\_address\_reviews} must contain exactly one review per item in \texttt{home\_address\_candidates}, in the same order.
    \item For each \(i\), \texttt{home\_address\_reviews[i].text} must equal \texttt{home\_address\_candidates[i]} exactly (character-for-character).
    \item If \texttt{home\_address\_candidates} is empty, \texttt{home\_address\_reviews} must be an empty list \texttt{[]}.
    \item \texttt{alphanumeric\_reviews} must contain exactly one review per item in \texttt{alphanumeric\_candidates}, in the same order.
    \item For each \(i\), \texttt{alphanumeric\_reviews[i].text} must equal \texttt{alphanumeric\_candidates[i]} exactly.
    \item If \texttt{alphanumeric\_candidates} is empty, \texttt{alphanumeric\_reviews} must be \texttt{[]}.
    \item Do not add extra reviews. Do not repeat a candidate. Do not output reviews for text that is not in the candidate lists.
    \item Never output an empty string as a candidate text.
\end{itemize}

\vspace{0.8mm}
\textbf{Hard rules:}
\begin{itemize}[leftmargin=5mm,itemsep=0.35mm,topsep=0.5mm]
    \item Do not invent any text not present in the narrative.
    \item For \texttt{KEEP} or \texttt{DROP}, you must include \texttt{evidence} copied verbatim from the narrative (short snippet).
    \item If you cannot find supporting evidence, mark \texttt{UNCERTAIN} and set \texttt{evidence} to \texttt{""}.
\end{itemize}

\vspace{0.8mm}
\textbf{Guidance:}
\begin{itemize}[leftmargin=5mm,itemsep=0.35mm,topsep=0.5mm]
    \item \texttt{HOME ADDRESS}: keep only the true residence or mailing address of a person. Drop crash-location addresses such as intersections, highways, mile markers, and scene locations.
    \item \texttt{ALPHANUMERIC IDENTIFIERS}: keep only personal identifiers such as license plates, driver's license or ID numbers, and SSNs. Drop roadway IDs (e.g., \texttt{I-94}, \texttt{US-12}), report or case numbers, incident IDs, tag numbers, and unit numbers unless clearly tied to a personal identifier.
\end{itemize}

\vspace{0.8mm}
Output JSON only, matching the schema exactly.
}
\end{tcolorbox}
}
\end{tcolorbox}

\caption{\textbf{Structured verifier system prompt used for candidate reviewing.}
}
\label{fig:verifier_system_prompt}
\end{figure*}

Operationally, we treat the verifier as a precision gate on the two most ambiguous categories, including home addresses and alphanumeric identifiers. Items marked DROP are removed from the verified output, while KEEP items are retained. UNCERTAIN items can be retained or excluded depending on deployment preference (precision-first vs. recall-first). In the current implementation, we use a recall-oriented setting, where UNCERTAIN items are retained in the final output. Importantly, this stage creates an auditable trail (decision and evidence) that supports error analysis and governance. An output example of the verifier is shown in Figure~\ref{fig:verifier_review_example}.

\begin{figure}
\centering

\begin{tcolorbox}[
    width=0.96\columnwidth,
    colback=gray!3,
    colframe=black!25,
    boxrule=0.5pt,
    arc=2mm,
    left=2mm,right=2mm,top=1.5mm,bottom=1.5mm
]
{\small
\textbf{Verifier output example}

\vspace{1mm}

\begin{tcolorbox}[
    colback=white,
    colframe=black!20,
    boxrule=0.4pt,
    arc=1.5mm,
    left=1.5mm,right=1.5mm,top=1mm,bottom=1mm
]
{\footnotesize
\texttt{\textbf{verifier\_review}} \\[0.5mm]

\begin{tabularx}{\linewidth}{@{}>{\raggedright\arraybackslash}p{0.22\linewidth} X@{}}
\texttt{field}     & \texttt{home\_address\_reviews[1]} \\[0.8mm]
\texttt{text}      & \texttt{"4647 HIGHWAY 47"} \\[0.8mm]
\texttt{decision}  & \texttt{"DROP"} \\[0.8mm]
\texttt{reason}    & \texttt{"Crash location address, not a true residence/mailing address of a person."} \\[0.8mm]
\texttt{evidence}  & \texttt{"UNIT 1 HIT THE DRIVEWAY OF 4647 HIGHWAY 47."}
\end{tabularx}
}
\end{tcolorbox}
}
\end{tcolorbox}

\caption{\textbf{Example of verifier-generated review for a candidate home address.} 
The verifier rejects the string \texttt{"4647 HIGHWAY 47"} because the narrative evidence indicates that it refers to the crash location rather than a person's true residential or mailing address.}
\label{fig:verifier_review_example}
\end{figure}

%
\subsection{Baseline}
To benchmark the proposed agentic workflow, we consolidate all extraction approaches evaluated in our prior system into a unified baseline family. These baselines operate in a single-pass manner: they directly produce PII candidates (via rules or LLM tagging), without any downstream processing, verification or agent-driven error correction. This baseline family includes (i) a rule-based extractor (Presidio), (ii) prompt engineering LLM extraction, and (iii) domain-adapted fine-tuned LLM extraction.
\subsubsection{Presidio}
We adopt Microsoft Presidio as a reproducible off-the-shelf baseline for detecting structured PII in unstructured text. In our implementation, we use Presidio’s default AnalyzerEngine and configure it to detect six predefined entity types: "PERSON", "PHONE\_NUMBER", "EMAIL\_ADDRESS", "MEDICAL\_LICENSE", "US\_DRIVER\_LICENSE", and "US\_SSN". These attributes were selected based on their relevance to transportation crash reports and alignment with our defined PII categories. "MEDICAL\_LICENSE", "US\_DRIVER\_LICENSE", and "US\_SSN" are classified as alphanumeric characters. Presidio also supports “LOCATION”, but because our task definition treats home address as PII while explicitly excluding crash locations, we do not evaluate Presidio on the home address category in the baseline. 
\subsubsection{LLM with Prompt-based Few-shot Learning}
To avoid the privacy and cost constraints of cloud LLMs, we evaluate locally hosted open-source LLMs, including Llama 3.1-8B and Gemma 27B~\citep{team2024gemma}, as prompt-based baselines. We use a few-shot prompting strategy where the prompt includes explicit role instructions category definitions, tagging rules, and in-domain examples to guide the model’s tagging behavior. The small Llama 3.1-8B model exhibited limitations in generalization, so we also experimented with a larger model – Gemma 27B to improve detection accuracy. This model demonstrated substantially better performance, particularly in clarifying ambiguous terms and reducing hallucinations. However, the trade-off was a significant increase in processing time and system resource requirements, which poses practical constraints for real-time or large-scale deployment.
\subsubsection{Fine-Tuning LLM}
Because crash narratives contain domain-specific phrasing and ambiguity that can break general-purpose PII detectors, we observed limitations in using the base Llama 3.1-8B model. To address these challenges, we fine-tuned the model using Low-Rank Adaptation (LoRA)~\citep{hu2022lora}. Meanwhile, to improve the poor training performance caused by domain mismatch, we manually annotated a subset of crash narratives drawn from our larger corpus, using our custom-defined PII categories and specific tagging rules. Although this manual process was time-consuming, it provides the domain-relevant supervision necessary for effective fine-tuning. The fine-tuned Llama 3.1-8B model was trained with 2,000 manually labeled crash narratives for 1 epoch, using a single L4 GPU on Google Colab.

Overall, the baseline family spans three representative paradigms for PII detection in crash narratives. While this setting provides an appropriate benchmark, our experiments and error analysis indicate that performance is uneven across categories. Structured PII such as phone numbers and emails is reliably detected by pattern-based rules, whereas home address and alphanumeric identifiers remain challenging due to context ambiguity. These limitations motivate the discussion we had in Section 3.2. Our proposed agentic workflow moves beyond a single extractor toward a specialized, tool-integrated and auditable workflow.
\subsection{Evaluation Metrics}

Given that PII detection is inherently a classification task, this study adopts standard classification metrics, including accuracy, precision, recall, and F1-score, to assess and compare the performance of different models. To ensure a comprehensive evaluation that aligns with the multi-faceted nature of PII extraction in crash narratives, these metrics are computed at multiple levels of granularity.

At the narrative level, we evaluate whether each crash narrative contains any PII. Specifically, for a dataset of $N$ crash narratives, each narrative is assigned a binary ground truth label: 1 if it contains at least one instance of PII, and 0 otherwise. A model $f$ then produces a binary prediction for each narrative, indicating whether it detects any PII. Using these predictions and the ground truth, we compute precision, recall, accuracy and F1-score at the narrative level, providing a high-level assessment of a model’s ability to detect the presence of sensitive information.

At a finer resolution, we conduct a per-type evaluation for each specific PII category: names, home addresses, phone numbers, email addresses, and alphanumeric identifiers. For each PII type $i$, we count the total number of true instances, denoted as $n_i$, across the $N$ narratives. Given a model $f$, we also count the number of predicted instances of type $i$, denoted as $\hat{n_i}$. By comparing the predicted instances to the ground truth, we identify true positives (correctly predicted instances), false positives (incorrectly predicted instances), and false negatives (missed instances). These quantities are used to compute the precision, recall, and F1-score for each PII type individually.

For example, consider a scenario where there are 10 true name entities across 100 crash narratives. If a model predicts 5 name entities, of which 3 are correct, then the model achieves a precision of 0.6, recall of 0.3, and F1-score of 0.4. This per-type analysis allows for a nuanced understanding of how different models perform across various categories of PII, which is essential for real-world deployment where different types of information may carry different levels of sensitivity and risk.

By combining both narrative-level and per-type evaluations, we aim to capture both the broad effectiveness and the fine-grained strengths and weaknesses of each model, facilitating informed decisions about model selection and deployment in large-scale crash narrative datasets.

\section{Results and Discussions}
We evaluate the proposed system under the same test setting used in the baseline experiments and compare it with the previously reported approaches, including Presidio, prompt-based LLMs, and the fine-tuned LLM. In the revised workflow, we further report results for the Hybrid Extractor and the ensemble and verifier configuration. The Hybrid Extractor combines Presidio (for phone numbers and emails) with the fine-tuned LLM (for names, home addresses, and alphanumeric identifiers), while the verifier configuration adds an agentic reviewing stage for the two most ambiguous categories: home address and alphanumeric identifiers.
\begin{table}[H]
\centering
\caption{Evaluation Results for PII Detection Models}
\label{tab:pii_eval}
\begin{adjustbox}{width=\textwidth}
\begin{tabular}{llcccc}
\toprule
\textbf{Models} & \textbf{PII Category} & \textbf{Precision} & \textbf{Recall} & \textbf{F1-score} & \textbf{Accuracy} \\
\midrule

\multirow{6}{*}{\parbox{3.5cm}{\textit{1. Presidio}}} 
 & Name & 0.51 & 0.68 & 0.58 & - \\
 & Phone Number & 0.88 & 0.94 & 0.91 & - \\
 & Email & 1.00 & 1.00 & 1.00 & - \\
 & Alphanumeric & 0.08 & 0.39 & 0.14 & - \\
 & Home address & - & - & - & - \\
 & \textbf{Overall} & \textbf{0.38} & \textbf{0.93} & \textbf{0.54} & \textbf{0.78} \\
\midrule

\multirow{6}{*}{\parbox{3.5cm}{\textit{2. Llama3.1-8B}\\\textit{prompting}}} 
 & Name & 0.40 & 0.58 & 0.47 & - \\
 & Phone Number & 0.00 & 0.00 & 0.00 & - \\
 & Email & 0.50 & 0.33 & 0.40 & - \\
 & Alphanumeric & 1.00 & 0.05 & 0.10 & - \\
 & Home address & 0.50 & 0.11 & 0.18 & - \\
 & \textbf{Overall} & \textbf{0.41} & \textbf{0.83} & \textbf{0.55} & \textbf{0.81} \\
\midrule

\multirow{6}{*}{\parbox{3.5cm}{\textit{3. Gemma-27B}\\\textit{prompting(few shot)}}} 
 & Name & 0.98 & 0.45 & 0.61 & - \\
 & Phone Number & 0.88 & 0.94 & 0.91 & - \\
 & Email & 1.00 & 1.00 & 1.00 & - \\
 & Alphanumeric & 0.34 & 0.89 & 0.49 & - \\
 & Home address & 0.61 & 0.89 & 0.72 & - \\
 & \textbf{Overall} & \textbf{0.78} & \textbf{0.65} & \textbf{0.71} & \textbf{0.93} \\
\midrule


\multirow{6}{*}{\parbox{3.5cm}{\textit{4. Fine-tuned LLM}\\\textit{(Llama3.1-8B)}}} 
 & Name & 0.94 & 0.70 & 0.80 & - \\
 & Phone Number & 0.54 & 0.44 & 0.48 & - \\
 & Email & 1.00 & 0.33 & 0.50 & - \\
 & Alphanumeric & 0.48 & 0.61 & 0.54 & - \\
 & Home address & 0.56 & 0.56 & 0.56 & - \\
 & \textbf{Overall} & \textbf{0.91} & \textbf{0.86} & \textbf{0.88} & \textbf{0.97} \\
\midrule

\multirow{6}{*}{\parbox{3.5cm}{\textit{5. Agentic Workflow}}}
 & Name & 0.93 & 0.76 & 0.84 & - \\
 & Phone Number & 0.88 & 0.94 & 0.91 & - \\
 & Email & 1.00 & 1.00 & 1.00 & - \\
 & Alphanumeric & 0.54 & 0.75 & 0.63 & - \\
 & Home address & 1.00 & 0.64 & 0.78 & - \\
 & \textbf{Overall} & \textbf{0.82} & \textbf{0.94} & \textbf{0.87} & \textbf{0.96} \\
\bottomrule
\end{tabular}
\end{adjustbox}
\end{table}

\subsection{Overall performance}
Table \ref{tab:pii_eval} demonstrates the performance of all methods in detecting PII across categories. Results indicate that rule-based Presidio is recall-oriented but prone to false positives, resulting in relatively low precision. Moreover, model size has a significant impact on LLM performance. With prompt engineering, a large model (i.e., Gemma 27B) could achieve much better results compare to a relatively small model (i.e., Llama 8B). However, a small model can be substantially enhanced with fine-tuning. For example, fine-tuning increases the precision of Llama 8B from 0.41 to 0.91, outperforming Gemma 27B with prompting engineering. This result indicates the benefits of domain-specific adaptation. With only 2,000 training samples, a small model can achieve better results than a large model, leading performance gains while saving computing resources.

Table \ref{tab:pii_eval} also indicates that Presidio excels at detecting structured PII while LLMs perform better in detecting context-depedent PII. By routing different PII types to the most suitable models, the agentic workflow enhances overall performance while avoiding re-developing new models. Moreover, with the ensemble learning scheme and the verifier agent, the proposed workflow achieves the best and most balanced overall performance. The result also indicates 
a clear recall-oriented tradeoff. The agentic workflow is particularly useful when missing sensitive entities is more costly than reviewing additional false positives. As shown in the per-category analysis below, this tradeoff is concentrated in the most ambiguous categories (i.e., home address and alphanumeric identifiers), which are the primary targets of the verifier agent.




\subsection{Per-category performance}

\subsubsection{Context-Dependent PII}
\textit{Name}: The agentic workflow achieves strong name detection performance (Precision = 0.93, Recall = 0.76, F1 = 0.84), improving recall over other methods while maintaining comparable precision. In practice, the hybrid framework preserves the strong name-detection capability of the fine-tuned model while integrating complementary tools for other PII categories.

\textit{Home Address}: Home address remains one of the most challenging categories because many street strings in crash narratives are not PII. Models must rely on detailed context clues to distinguish residential addresses from non-residential ones, which requires strong contextual understanding. In the baseline family, Gemma-27B prompting achieves the strongest recall (0.89), but with moderate precision (0.61). The agentic workflow improves precision to 1.00 but reduces recall to 0.64, reflecting a stricter extraction behavior.

\textit{Alphanumeric identifiers}: Alphanumeric identifiers are similarly challenging because many non-PII tokens in crash narratives share similar surface forms (e.g., roadway IDs, report numbers, unit references). The agentic workflow reaches a more balanced result than several baselines (Precision = 0.54, Recall = 0.75, F1 = 0.63), but this category continues to generate both false positives and false negatives.

\subsubsection{Structured PII}
\textit{Phone Number}: Presidio achieves a very good and balanced performance with a high precision of 0.88 and a high recall of 0.94. This confirms that phone numbers are a structure-dominant category and are better handled by deterministic rule-based extraction. The agentic workflow inherits this strength directly from Presidio, and therefore matches the same strong performance. 

\textit{Email Address}: Email addresses are also highly structured and benefit from rule-based extraction. Presidio performed perfectly with a 1.00 F1 score. Therefore, our proposed agentic workflow achieved the same strong performance.

Overall, these category-level results reinforce the central motivation for the hybrid design: structured PII should be handled by rules, while contextual PII requires a trained language model. However, home addresses and alphanumeric identifiers still show substantial residual error, motivating the agentic verifying stage.

\subsection{Ablation Study}
The module of ensemble learning and verifier is designed to address the challenges in detecting home addresses and alphanumeric identifiers. The workflow applies an ensemble over multiple fine-tuned LLM runs to improve candidate recall, followed by a verifier agent that reviews candidate spans and filters likely false positives. We report this agentic workflow only for home address and alphanumeric identifiers, since the verifier is intentionally scoped to those high-risk categories and does not modify name, phone number, or email outputs. 

To further measure the module influence on the workflow, we conduct an ablation study, with results presented in Table~\ref{tab:tp_fp_fn}. For the home address category, adding the module substantially improves performance, increasing precision, recall, and F1 from 0.71/0.45/0.56 to 1.00/0.64/0.78. The TP/FP/FN counts in Table~\ref{tab:tp_fp_fn} provide a clearer view of this gain: true positives increase from 5 to 7, false positives decrease from 2 to 0, and false negatives decrease from 6 to 4. This represents a substantial improvement for the most context-sensitive category. In practice, the verifier helps reject crash locations and roadway references that superficially resemble residential addresses, while the ensemble stage recovers valid home addresses that may be missed in a single run of the fine-tuned tagger. Together, these components yield a much better balance between precision and recall.


For alphanumeric identifiers, adding the module improves recall and F1, changing precision/recall/F1 from 0.56/0.65/0.60 to 0.54/0.75/0.63. As shown in Table~\ref{tab:tp_fp_fn}, this improvement is primarily recall-driven: true positives increase from 13 to 15 and false negatives decrease from 7 to 5, although false positives also increase from 10 to 13. This suggests that the ensemble stage expands candidate coverage and recovers identifiers missed in a single run, while the verifier removes only part of the additional noise. Alphanumeric strings remain inherently difficult to disambiguate because many non-PII tokens share formats similar to true identifiers. As a result, adding the module yields a modest F1 improvement, but unlike home-address detection, the gain is driven mainly by higher recall rather than improved precision.


\begin{table}[H]
\centering
\caption{Ablation of the ensemble learning + verifier (E+V) module on the two most ambiguous categories.}
\label{tab:tp_fp_fn}

\setlength{\tabcolsep}{5pt}
\renewcommand{\arraystretch}{1.1}

\begin{tabular}{l|cc|cc}
\toprule
 & \multicolumn{2}{c|}{Home Address} & \multicolumn{2}{c}{Alphanumeric Identifier} \\
 & w/o E+V & w/ E+V & w/o E+V & w/ E+V \\
\midrule
TP (↑) & 5 & 7 & 13 & 15 \\
FP (↓) & 2 & 0 & 10 & 13 \\
FN (↓) & 6 & 4 & 7 & 5 \\
\bottomrule
\end{tabular}

\end{table}





\subsection{Discussions}
The experimental results demonstrate the effectiveness of the proposed agentic workflow for detecting PII. First, the Hybrid Extractor improves end-to-end detection by combining the strengths of Presidio and the fine-tuned LLM. Second, the Verifier further improves workflow performance by enforcing a second-pass contextual judgment with explicit evidence requirements, which acts as a precision control mechanism. Third, the ensemble-based candidate generation introduces an important recall–precision trade-off. By collecting candidates across multiple runs of the fine-tuned LLM, the system captures variability that would otherwise cause missed entities in a single decoding pass. However, this also expands the candidate pool and can introduce additional false positives, especially for alphanumeric identifiers. The verifier mitigates this effect but does not fully eliminate it, suggesting an opportunity for future work on stronger verifier prompts, majority-vote auditing, or category-specific verifier policies.

These findings support the broader claim of this paper. For domain-specific PII detection in crash narratives, a purely monolithic extractor is suboptimal. A tool-integrated, agentic workflow with a specialized extractor and context-aware verifier provides a more reliable and deployable approach. 

\section{Conclusions}
This study addresses the practical challenges in detecting PII in crash narratives, where sensitive information appears in highly variable and context-dependent forms. Rather than relying on a single extractor, we formulate the task as a tool-integrated workflow and show that different PII categories benefit from different extraction mechanisms. In particular, structured PII is reliably handled by rule-based extraction, while context-dependent PII is better handled by a domain-adapted fine-tuned LLM.

Based on this observation, we propose a hybrid, agentic workflow that combines category-specific extraction with targeted verification. The Hybrid Extractor routes structured PII to Presidio and contextual PII to a fine-tuned Llama 3.1-8B model, improving system-level performance through orchestration rather than changes to the underlying base model. We further introduce a module of ensemble learning and verification for home addresses and alphanumeric identifiers, the two most error-prone categories, and show that this stage improves robustness, particularly for home address detection.

A key practical advantage of the proposed system is that all components are deployed locally, making the system compatible with privacy-preserving operational environments where external API usage is restricted. Overall, the results support a tool-integrated and auditable design for domain-specific PII detection, and suggest that future improvements should focus on stronger verifier policies and category-specific uncertainty handling for ambiguous PII types.

\section{Acknowledgments}
This research is supported by the Wisconsin Traffic Records Coordinating Committee (TRCC). The ideas and views expressed in this paper are those of the authors and do not necessarily reflect the views of TRCC.


\bibliographystyle{cas-model2-names}

\bibliography{cas-refs}

\end{document}